\documentclass[%
 aip,
 jap,%
 amsmath,amssymb,amsfont,
 draft
 preprint,%
]{revtex4-1}

\usepackage{graphicx}
\usepackage{bm}

\begin{document}

\title{Influence of cross-section geometry and wire orientation on the phonon shifts in ultra-scaled Si nanowires}%

\author{Abhijeet Paul}
\email{abhijeet.rama@gmail.com}
\author{Mathieu Luisier}
\author{Gerhard Klimeck}

\affiliation{School of Electrical and Computer Engineering, Network for Computational Nanotechnology, \\ Purdue University, West Lafayette, Indiana, USA, 47907.}

\date{\today}

\begin{abstract}

Engineering of the cross-section shape and size of ultra-scaled Si nanowires (SiNWs) provides an attractive way for tuning their structural properties. The acoustic and optical phonon shifts of the free-standing circular, hexagonal, square and triangular SiNWs are calculated using a Modified Valence Force Field (MVFF) model. The acoustic phonon blue shift (acoustic hardening) and the optical phonon red shift (optical softening) show a strong dependence on the cross-section shape and size of the SiNWs. The triangular SiNWs have the least structural symmetry as revealed by the splitting of the degenerate flexural phonon modes and The show the minimum acoustic hardening and the maximum optical hardening. The acoustic hardening, in all SiNWs, is attributed to the decreasing difference in the vibrational energy distribution between the inner and the surface atoms with decreasing cross-section size. The optical softening is attributed to the reduced phonon group velocity and the localization of the vibrational energy density on the inner atoms. While the acoustic phonon shift shows a strong wire orientation dependence, the optical phonon softening is independent of wire orientation. 



\end{abstract}


\maketitle 
\section{Introduction} \label{sec:I}


Silicon nanowires (SiNWs) have the potential to become the next generation solution for improved (i) Complementary-Metal-Oxide-Semiconductor (CMOS) transistors with higher channel drive currents \cite{Sinwfet}, (ii) thermoelectric (TE) devices \cite{Hochbaum2008}, (iii) rechargeable batteries \cite{sinw_battery}, (iv) compact non-volatile memories \cite{sinw_nvm}, (v) solar cells, \cite{sinw_solar_cells} and (vi) explosive and bio-sensors \cite{sinw_explosive_sensor,sinw_biosensor}. 
Extensive application of SiNWs to various technologies requires to understand the structural \cite{SINW_110_phonon}, electrical\cite{anantram_ph}, thermal\cite{Mingo_kappa} and optical \cite{sinw_solar_cells} properties which arise from the strong geometrical confinement, atomic positions, and increased surface-to-volume ratio (SVR). Especially the phonon spectra of SiNWs can provide a lot of vital information about their structural properties which can be optimized for maximum technological benefits in SiNWs.

Engineering the Si nanowire structural properties based on their cross-section size has been studied extensively \cite{exp_opt_phonon_sinano,exp_ac_phonon,mahan_111,hepplestone_sinw_2}. However, the impact of cross-section shape has not been completely addressed yet. Experimentally, the impact of semiconductor (GaAs, GaP, ZnS, etc) nanowire shape on their structural properties  has been observed \cite{shape_dep_raman,shape_dep_1,shape_eff_phd}.  Motivated by these experiments, we theoretically explore the effect of the cross-section geometry of ultra-scaled SiNWs on their structural properties like phonon shifts. We furthermore provide analytical expressions for the variation of these physical quantities with size and shape thus, enabling (i) physical insights, and (ii) a compact modeling representation of the structural properties.


The determination of size and stability of SiNWs is crucial for their usage in different applications \cite{Sinwfet,sinwfet2,Hochbaum2008,Boukai2008,sinw_battery,sinw_nvm,sinw_solar_cells,sinw_explosive_sensor,sinw_biosensor}. Raman spectroscopy provides a valuable, non-destructive method to analyze the physical properties of nanostructures (NS) like cross-section size \cite{exp_ac_phonon,exp_opt_phonon_sinano}, degree of crystallinity \cite{adv_RS_method}, dimensionality of confinement \cite{exp_opt_phonon_sinano,size_dep_red_shift}, etc. Raman spectroscopy utilizes the amount of phonon spectrum shift and line width broadening to obtain useful structural information \cite{arora_opt_phonon}.  

The first satisfactory theoretical model for explaining Raman shifts and line width broadening was proposed by Richter et al \cite{Richter} and improved by Campbell and Fauchet \cite{Campbell}. These phenomenological continuum models are based on the choice of the phonon weighting function as well as the boundary conditions of the phonon vibrations \cite{Richter,Campbell} which are chosen rather arbitrarily and hence do not provide correct phonon shift in nanostructures \cite{Zi_PRB}. Arora et al. experimentally \cite{arora_opt_phonon} reported that phonon shifts are very different from bulk in nanostructures with size less than 20nm. Tanaka et al \cite{Tanaka} also pointed out the failure of continuum models to predict the Raman shift in nanostructures with size $<$ 7.5nm due to the arbitrary parameter sets used in continuum models. While continuum models can match the experimental Raman spectroscopy data using fitting parameters, they lack the predictive capability for estimating the phonon shifts and line widths in nanostructures with different shapes and compositions \cite{Zi_PRB,Tanaka,raman_size_dep}.

The evident inability of the continuum models to properly explain and predict the structural properties in nanostructures \cite{Zi_PRB,Tanaka,raman_size_dep} demands for methods which can automatically take into account the effects of structural miniaturization like geometrical confinement, orientation effects, cross-sectional shape and surface-to-volume ratio effects. 



Many efforts to explain the phonon shifts in nanostructures have been reported. They are based on models like the VFF approach \cite{VFF_kanelis,mahan_111}, partial density approach \cite{Zi_PRB}, Bond Charge Model (BCM) \cite{hepplestone_sinw_1,hepplestone_sinw_2}, bond order model \cite{raman_size_dep}, etc. In all these approaches the importance of taking into account the atomic positions and the atomic surrounding to understand the Raman shift has been stressed which most of the continuum models lack \cite{Richter, Campbell}.

We utilize a Modified Valence Force Field (MVFF) phonon model \cite{VFF_mod_herman,jce_own_paper,iwce_own_paper} to study the physical properties of SiNWs. This MVFF model (i) is relatively straight forward to understand and implement, (ii) explains the experimental phonon and elastic properties of zinc-blende materials very well \cite{VFF_mod_herman,VFF_mod_zunger,jce_own_paper,VFF_kanelis}, and (iii) is computationally very efficient which allows for expanding the simulation domain easily to few million atoms \cite{anharmonic}. However, frozen phonon models like the MVFF one do not integrate thermodynamics which forbids the correct calculation of many physical properties such as low temperature volume thermal expansion. Yet these models give a good insight into the physical processes governed by the lattice under ambient conditions.

The paper is organized as follows. Section~\ref{sec:II} provides a brief introduction about the MVFF model for the calculation of phonons in SiNWs and about the method to extract the phonon shifts in these nanowires (Sec.~\ref{sec:wire_str}). The results section (Sec.~\ref{sec:Res}) discusses the effect of cross-section shape and size on the phonon spectra (Sec.~\ref{phon_disp}) and presents experimental benchmarking of the phonon shifts using the MVFF model (Sec. \ref{sec:exp_res}). Furthermore, the trends of the acoustic phonon shift (Sec.~\ref{sec:Res_1}) and the optical phonon shift (Sec.~\ref{sec:Res_2})
are discussed along with wire orientation dependence (Sec.\ref{sec:or_dep}). Conclusions are provided in Section~\ref{sec:conc}.

\section{Theory and Approach} 
\label{sec:II}
\subsection{Phonon model} \label{sec_phon_model}

In the MVFF model, the frequencies of the phonon modes are calculated from the forces acting on atoms induced by finite displacements of the atoms from their equilibrium positions in a crystal \cite{phon_rev_paper}. To calculate the restoring force(F), first the total potential energy of the solid (U) is estimated. For the MVFF model, U is approximated as \cite{jce_own_paper}, 

\begin{eqnarray}
\label{eq:uelastic}
 U & \approx & \frac{1}{2} \sum_{i\in N_{A}} \Bigg[ \sum_{j\in nn(i)} U^{ij}_{bs} + \sum^{j \neq k}_{j,k \in nn(i)} \big( U^{jik}_{bb} \nonumber \\
            & &  + U^{jik}_{bs-bs} + U^{jik}_{bs-bb} \big) + \sum^{j \neq k \neq l}_{j,k,l \in COP_{i}} U^{jikl}_{bb-bb} \Bigg],
\end{eqnarray}

where $N_{A}$, $nn(i)$ and $COP_{i}$ represent the total number of atoms in one unit cell, the number of nearest neighbors for atom `i', and the coplanar atom groups for atom `i', respectively. The first two terms $U^{ij}_{bs}$ and $U^{jik}_{bb}$ represent the elastic energy coming from bond stretching and bending between the atoms connected to each other as given in the original Keating VFF model \cite{Keating_VFF}. 
The terms $U^{jik}_{bs-bs}$, $U^{jik}_{bs-bb}$ and $U^{jikl}_{bb-bb}$ represent the cross bond stretching \cite{VFF_mod_herman}, cross bond bending-stretching \cite{VFF_mod_zunger}, and coplanar bond bending \cite{VFF_mod_herman} interactions, respectively. The detailed procedure for obtaining the phonon spectra in Si bulk and NWs are provided in Ref.\cite{jce_own_paper,iwce_own_paper}. In the following sections the method to extract the acoustic and optical phonon shifts and the types of SiNWs used in this study are discussed. 


\subsection{Phonon shift in Si nanowires}
\label{sec:wire_str}

The vibrational spectra of the nanostructures are strongly influenced by (i) the finite size of the nanostructures, (ii) the mismatch with the matrix material ( the surrounding atoms) or (iii) defects \cite{arora_opt_phonon,exp_ac_phonon,adv_RS_method}. The optical and the acoustic phonon energies are shifted compared to the bulk values (around $q \propto 0$) in these geometrically confined structures. Raman spectroscopy reveals the decrease of the optical frequency with down-scaling of the nanostructures (called the optical softening or optical red shift) \cite{raman_size_dep,jusserand} and the increase of the acoustic frequency (called the acoustic hardening or acoustic blue shift) \cite{exp_ac_phonon,raman_size_dep}.  These phonon shifts are estimated using the following relation \cite{raman_size_dep,hepplestone_sinw_1,hepplestone_sinw_2}, 
 

\begin{equation}
\label{phonon_shift}
	\Delta \omega^{ac/opt}_{0}  =  \omega^{ac/opt}_{NW}-\omega^{ac/opt}_{bulk},
\end{equation}
where $\omega^{ac/opt}_{bulk}$ is the bulk acoustic (optical) phonon frequency at the $\Gamma$ point. The phonon shifts show the following functional dependence on the nanostructure cross-section size (W) \cite{raman_size_dep,exp_opt_phonon_sinano,hepplestone_sinw_1}, 

\begin{equation}
\label{phonon_shift_size_dep}
	\Delta \omega^{ac/opt}_{0} = A_{ac/opt}(\frac{a_{0}}{W})^{d_{ac/opt}}\;\;[cm^{-1}],
\end{equation}
where $a_{0}$ is the lattice constant of the material. For  A$>$0 the phonon shift is positive (or blue shift) and for A$<$0 the shift is negative (or red shift) \cite{raman_size_dep}. The exponent value `d' indicates the dimensionality of the geometrical confinement. For 1D nanowires/nanorods a variety of `d' values have been reported in the literature like 0.67 \cite{hepplestone_sinw_1}, 0.95 \cite{SINW_110_phonon}, 1.44\cite{exp_opt_phonon_sinano}, etc. 

Using a series of Raman spectroscopy measurement for a sample of nanostructures the size and the dimensionality of the nanostructures can be predicted \cite{arora_opt_phonon,raman_size_dep,shape_dep_raman}. The phonon shifts observed in Raman spectroscopy can be predicted using the phonon spectrum. Atomistic phonon modeling correctly takes surface boundary conditions into account depending on the cross-section shape of the wire which reflects in the overall phonon spectrum. The surface and inner atomic movements vary with different cross-section shapes and therefore affects the phonon shifts. Equation \ref{phonon_shift_size_dep} captures the nanostructure size effect in the exponent value `d' and the shape effect in the pre-factor `A'. Thus, a combination of A and d can be useful for the determination of size and shape of the SiNWs.


\subsection{Si Nanowire details}
\label{sinw_detail}
\begin{figure}[!hbt]
	\centering
		\includegraphics[width=2.9in,height=2.4in]{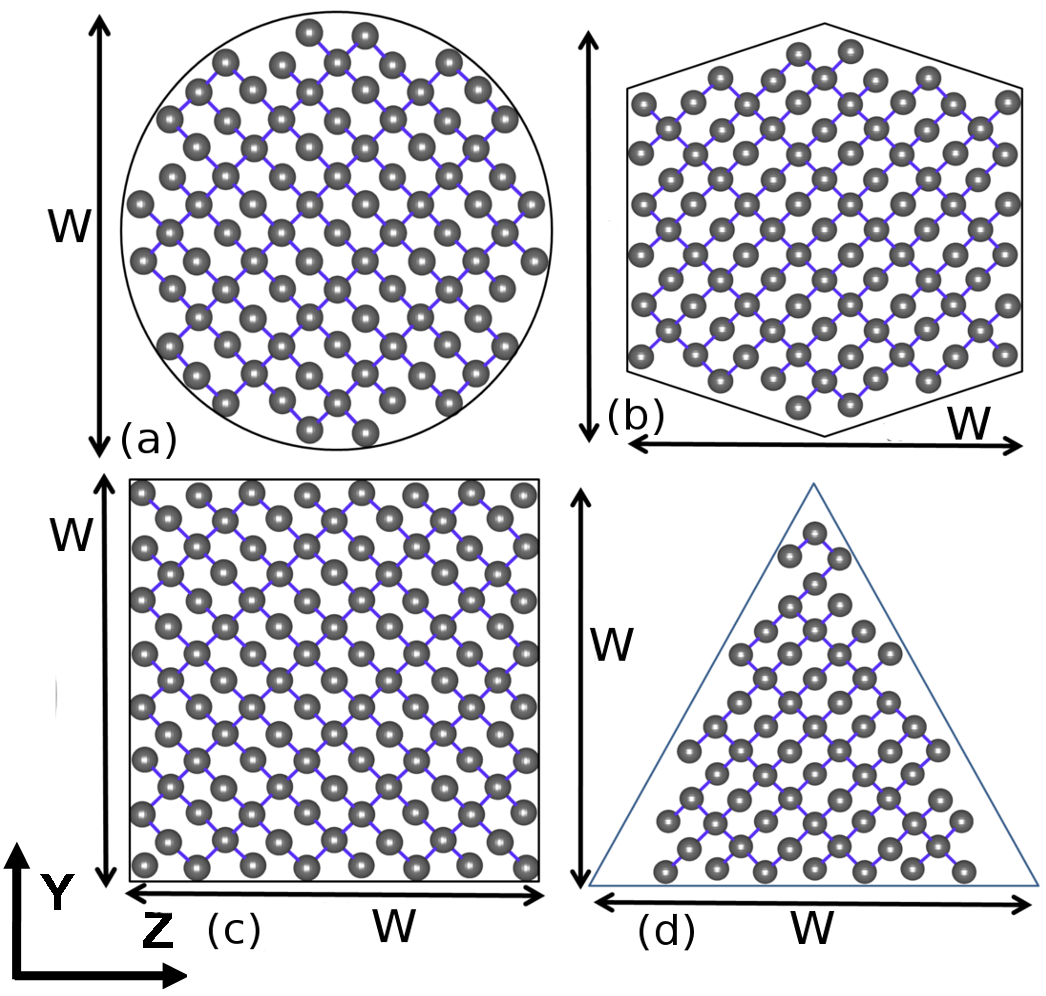}
	\caption{Projected unit cell structures of free-standing $[$100$]$ oriented silicon nanowires with (a) Circular, (b) Hexagonal, (c) Square, and (d) Triangular cross-section shapes. Width and height of the cross-section are defined using a single width variable W (width = height). These structures are at W = 2nm.}
	\label{fig:SiNW_unitcell}
\end{figure}

Four cross-section shapes for $[$100$]$ oriented SiNWs have been considered in this study, (a) circular, (b) hexagonal, (c) square and (d) triangular cross-section shapes (Fig.~\ref{fig:SiNW_unitcell}). SiNWs with [110] and [111] channel orientation studied here are of square cross-section only. The feature size is determined by the width parameter W. The value of W is varied from 2 to 6 nm. The surface atoms are allowed to vibrate freely without any passivating species. The wires are still assumed to have a tetrahedral geometry. It has been shown that wires with diameter below 2nm tend to lose the tetrahedral structure \cite{sinw_stability,Amrit} due to surface pressure and internal strain.


\section{Results and Discussion} 
\label{sec:Res}

In this section the effect of cross-section on the phonon dispersion and phonon shifts of $[$100$]$ SiNWs are discussed.There is a lack of proper channel orientation information in the literature.  However Adu et al. \cite{shape_dep_shift_exp} reported $[$100$]$ channel orientation due to which we chose [100] SiNWs for shape study. Furthermore, we also investigate the impact of channel orientation on the phonon shifts in square SiNWs.

\subsection{Phonon dispersion}
\label{phon_disp}

\begin{figure}[htb!]
	\centering
		\includegraphics[width=3.0in,height=2.9in]{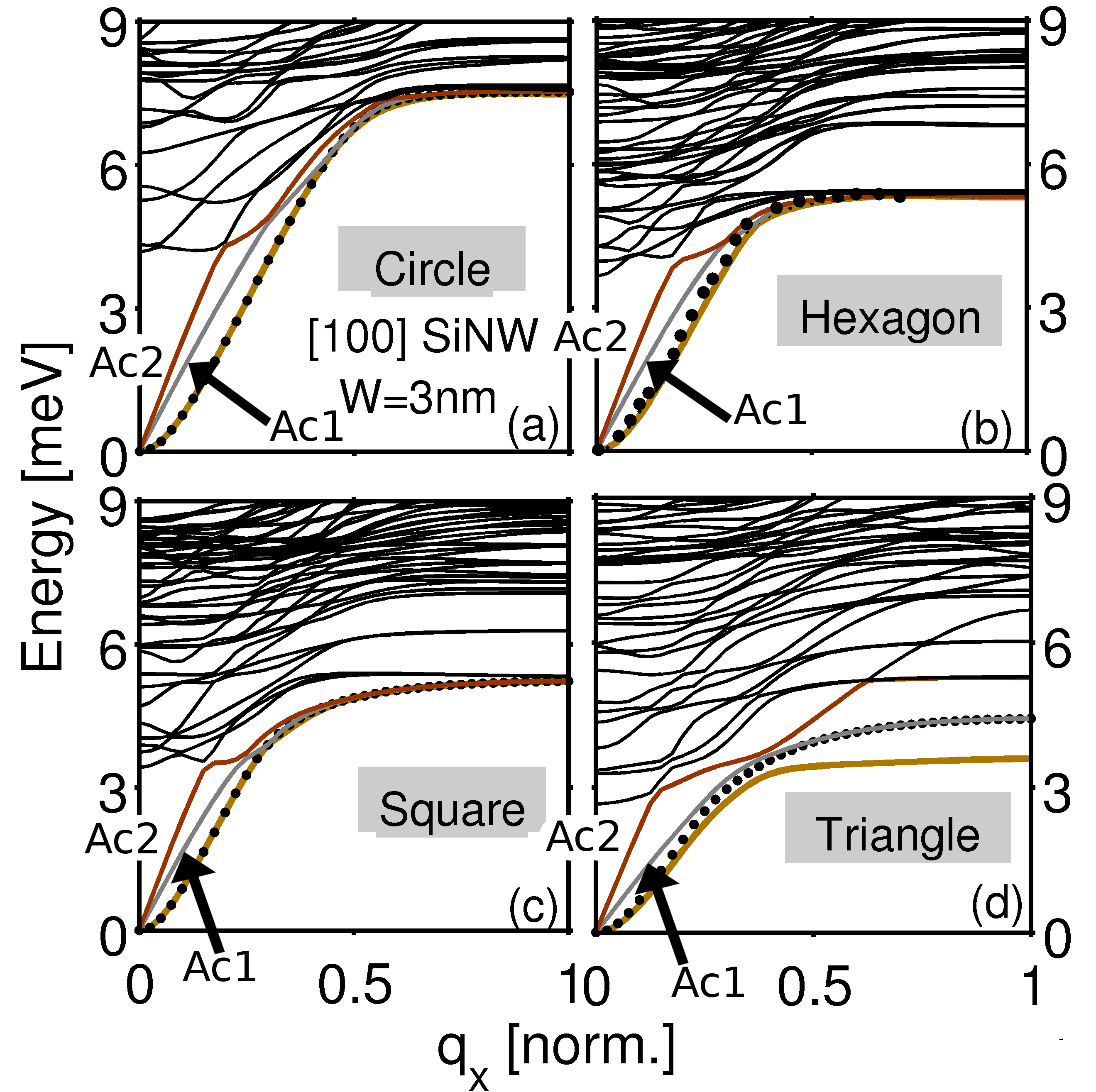}
	\caption{Phonon dispersion in free-standing $[$100$]$ oriented, W = 3nm, SiNWs with (a) Circular, (b) Hexagon, (c) Square, and (d) Triangular cross-section shapes. The first two branches (brown and black dot) are the flexural modes \cite{iwce_own_paper}. The next two branches (Ac1, Ac2) are the acoustic branches \cite{iwce_own_paper}. }
	\label{fig:SiNW_phonon}
\end{figure}

The phonon dispersion of SiNWs for all the cross-section shapes with $[$100$]$ channel are shown in Fig.~\ref{fig:SiNW_phonon}. While the full phonon dispersion extends up to $\sim$65meV, phonon modes till 9 meV energy range are shown for clarity. All the wires exhibit two flexural branches\cite{iwce_own_paper} ($\omega_{q\leftarrow0} \propto q^2$). These modes are double degenerate in all the shapes except the triangular wire. In the triangular wires these modes split due to the reduced structural symmetry (Fig.~\ref{fig:SiNW_phonon} d). The next two phonon branches in all the structures show $\omega_{q\leftarrow0} \propto q$. These branches determine the sound velocity in these structures \cite{iwce_own_paper} (Table \ref{table_0}). The triangular wire has the lowest longitudinal and transverse sound velocity ($V_{snd,l}$) whereas other shapes have very similar sound velocities. \textit{The splitting of the degenerate flexural modes and the reduced sound velocity show that triangular wires have the least structural symmetry.}

\subsection{Experimental Benchmark}
\label{sec:exp_res}

Before using the MVFF model to calculate phonon frequency shifts it is important to compare the theoretical MVFF results with experimental data. 
Both the experimental acoustic and optical phonon shifts in Si nanocrystals are compared. 

\begin{figure}[htb!]
	\centering
		\includegraphics[width=2.5in,height=2.4in]{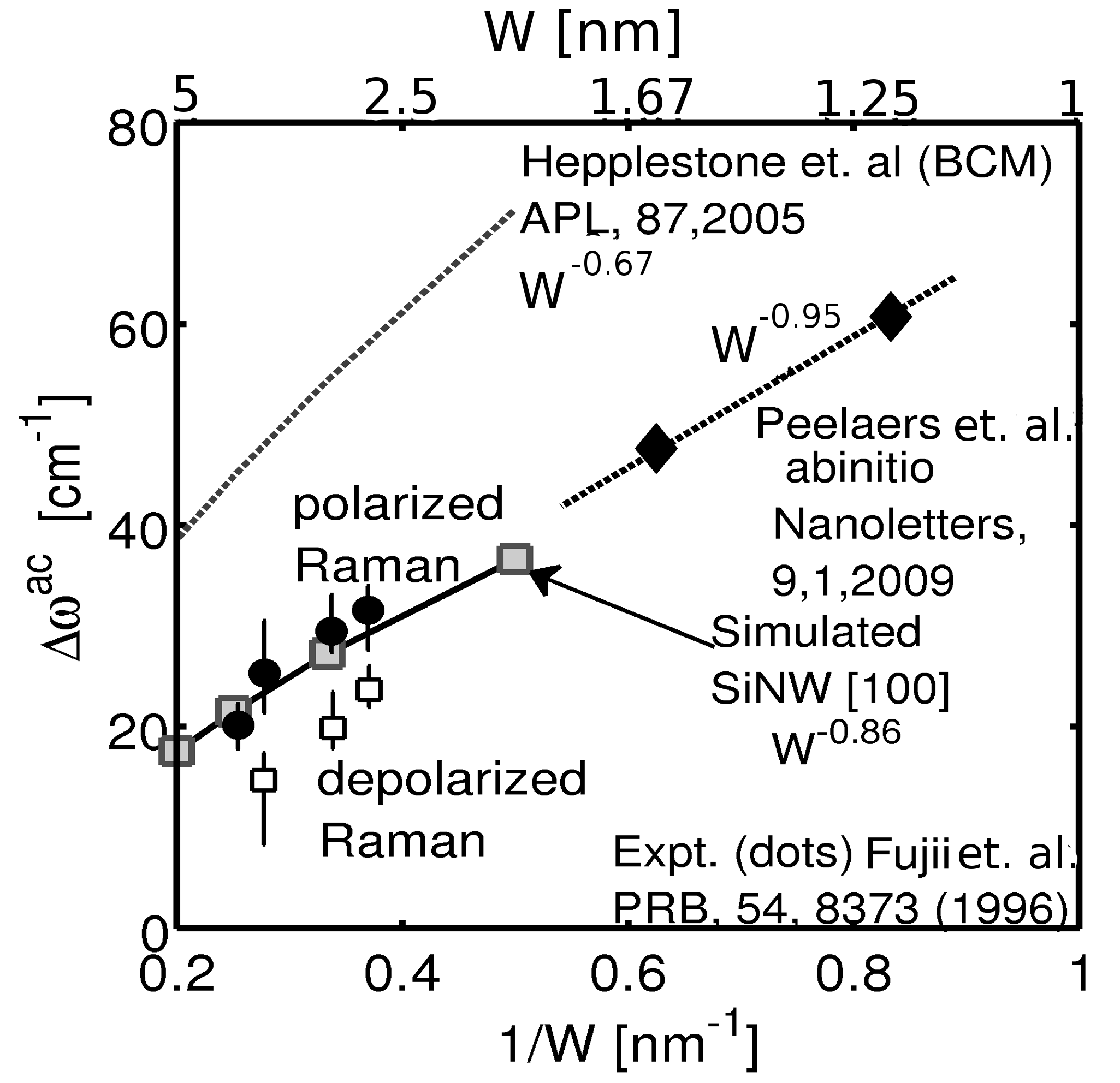}
	\caption{Comparison of theoretical acoustic phonon frequency shift (line) obtained for circular $[$100$]$ SiNWs using MVFF model with experimental phonon shift data (dots with error bars) \cite{exp_ac_phonon}. The diamond dots show the abinitio based results for $[$110$]$ SiNWs \cite{SINW_110_phonon}. The BCM based results \cite{hepplestone_sinw_2} are shown using a dotted line. }
	\label{fig:exp_act_benchmark}
\end{figure}

\textit{Acoustic shift:} The experimental values of acoustic phonon confinement (shown as dots with error bars in Fig.~\ref{fig:exp_act_benchmark}) using Raman spectroscopy for Si nanocrystals embedded in a $SiO_{2}$ matrix have been reported by Fujii et. al \cite{exp_ac_phonon}.  The calculated acoustic phonon shifts in similar size free-standing circular $[$100$]$ SiNWs using the MVFF phonon model compare quite well to the experimental data as shown in Fig. \ref{fig:exp_act_benchmark}, providing frequency shifts between 20 and 30 $cm^{-1}$. The MVFF calculated values show a power law of $W^{-0.86}$ close to the results for $[$110$]$ SiNWs (diamond dots in Fig. \ref{fig:exp_act_benchmark}) from Ref. \cite{SINW_110_phonon} ($W^{-0.95}$). The difference in the power laws may arise from the wire surface passivation, structural relaxation and channel orientation. The theoretical predictions using the BCM model \cite{hepplestone_sinw_2} gives a cross-section dependence of $W^{-0.67}$ which overestimates the experimentally observed acoustic phonon shift (Fig. \ref{fig:exp_act_benchmark}).

\begin{figure}[htb!]
	\centering
		\includegraphics[width=2.7in,height=2.2in]{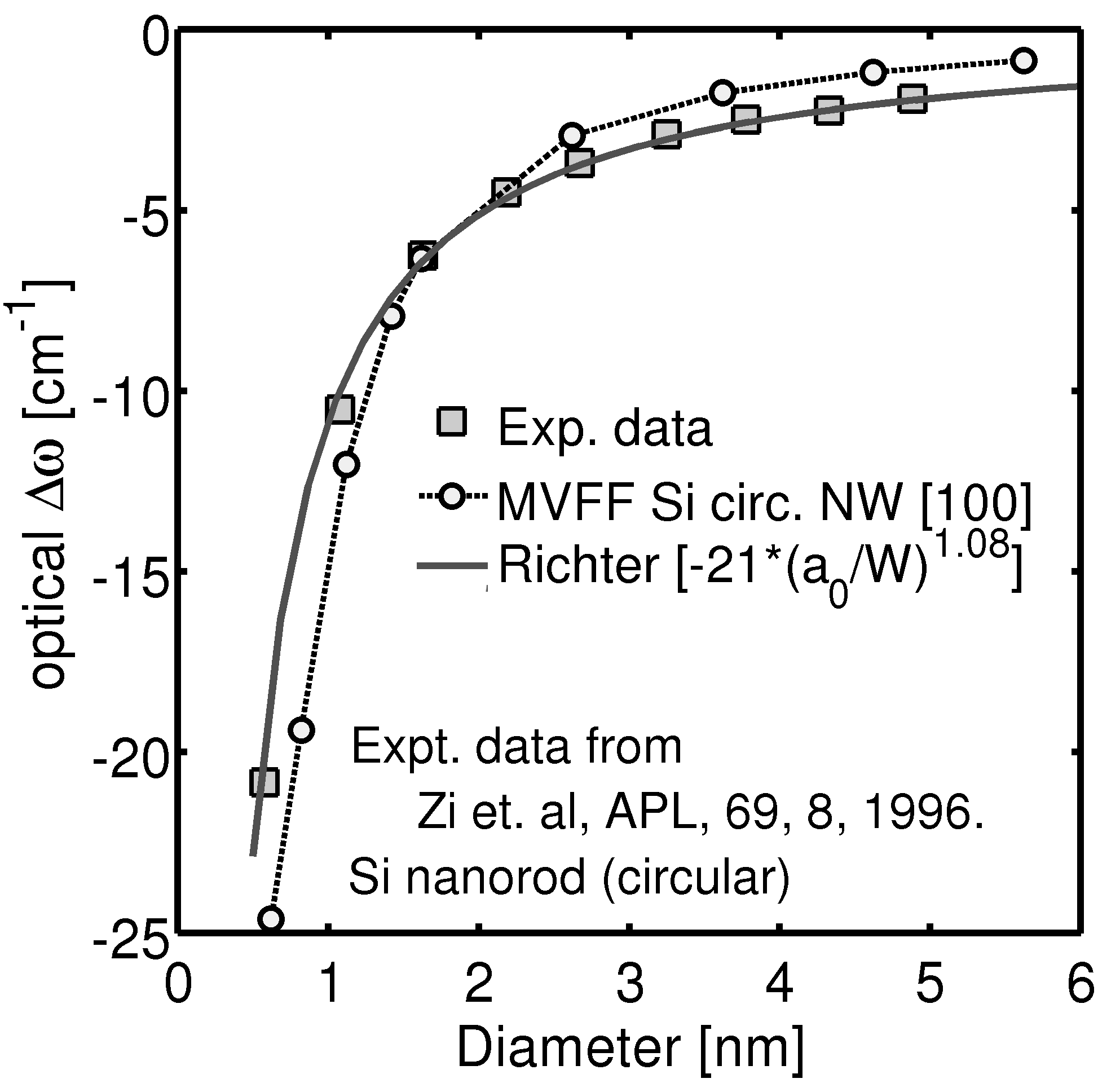}
	\caption{Comparison of theoretical optical phonon frequency shift (line), obtained for circular SiNWs using the MVFF model, and experimental data. The experimental optical phonon shift is for silicon nanorods embedded in $\text{SiO}_{2}$ \cite{exp_opt_phonon_sinano}. The pre-factor and the exponent used for Richter's model is obtained arbitrarily by matching the experimental data \cite{exp_opt_phonon_sinano}. Though the match is good the fitting procedure lacks any predictive capability contrary to the MVFF model.}
	\label{fig:exp_optbenchmark}
\end{figure}

\textit{Optical shift}: The optical phonon shift is experimentally measured using Raman spectroscopy \cite{exp_opt_phonon_sinano,shape_dep_raman}. Figure \ref{fig:exp_optbenchmark} shows the experimental optical phonon shift result for Si nanorods (symbols) embedded in a $SiO_{2}$ matrix. With decreasing cross-section size the optical red-shift increases \cite{raman_size_dep}. The theoretical optical phonon shift is calculated for free-standing $[$100$]$ circular SiNWs using the MVFF model. The theoretical values of the optical shift agree well with the experimental data (Fig. \ref{fig:exp_optbenchmark}). The mismatch in phonon shifts at extremely small dimensions ($W<2nm$) may be due to, (i) the structural relaxation in small cross-section wires \cite{sinw_stability,Amrit} which is not taken into account in this study and (ii) the presence of $SiO_{2}$ around SiNWs which might affect the results for smaller wires though the overlap of the optical phonon spectrum of these two materials is small \cite{arora_opt_phonon}. Hence, the MVFF model correctly captures the optical phonon shifts in SiNWs for W$\ge$2nm, however, for $W<2nm$ the MVFF phonon values without structural relaxation are not very accurate. Though the optical shift is very well captured by the Richter model \cite{Richter}, the fitting parameters are chosen arbitrarily which removes the predictive capability from the model.

In the next few sections we will discuss the structural properties of ultra-scaled SiNWs with different cross-section shapes and sizes.

\subsection{Acoustic Phonon shift in SiNWs}
\label{sec:Res_1}



\begin{figure}[!htb]
	\centering
		\includegraphics[width=3.4in,height=1.9in]{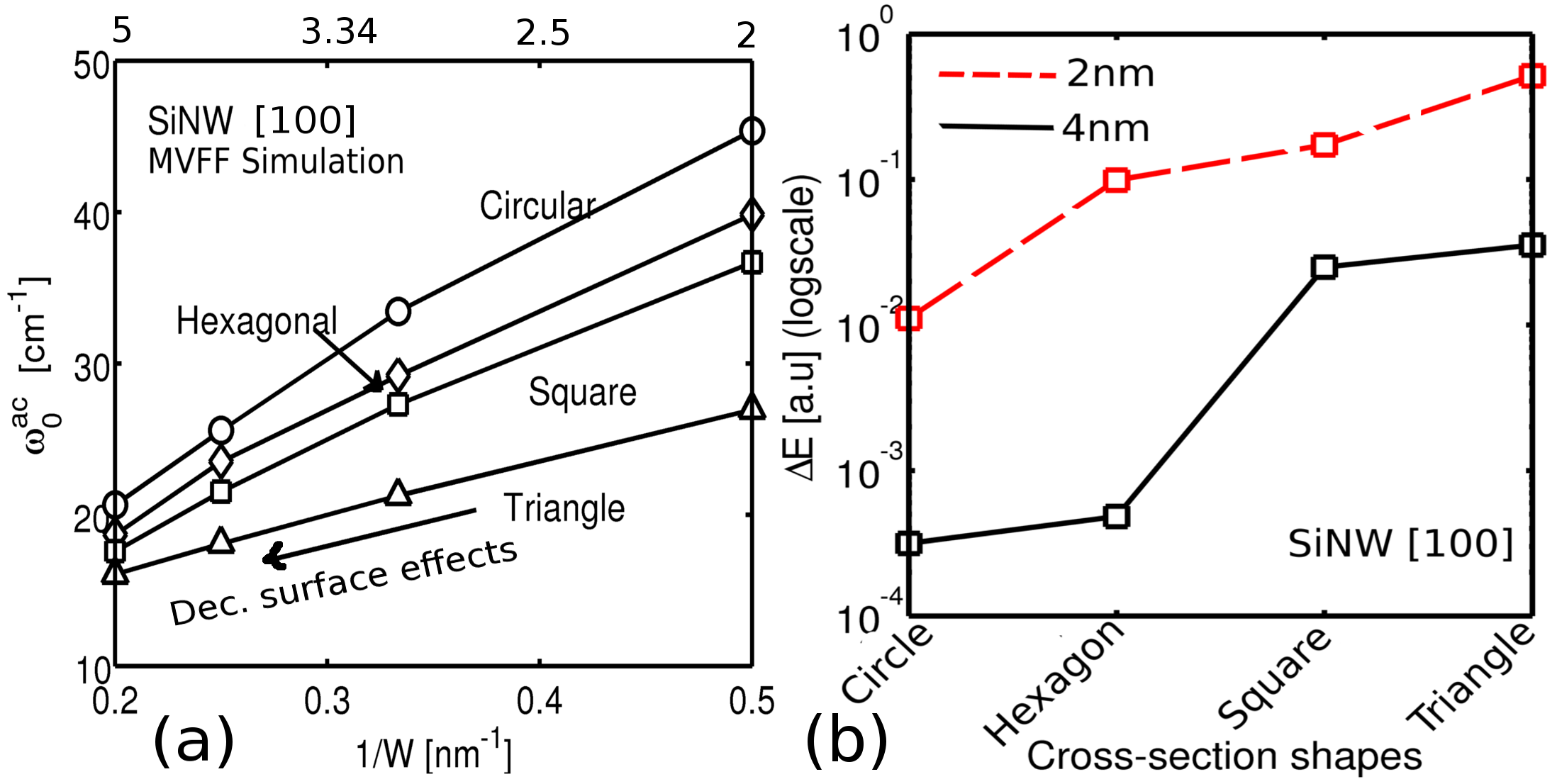} 
	\caption{(a) Effect of cross-section shape on the acoustic phonon shift in $[$100$]$ SiNWs. (b) Difference in the average vibrational energy density of the inner and the surface atoms in 4 cross-section shape [100] SiNWs for W = 2nm and 4nm. }
	\label{fig:ac_shift}
\end{figure}

\textit{Influence of cross-section shape and size:} The acoustic phonon shifts are calculated using the MVFF model for different cross-section shaped SiNWs. All the wire geometries show a blue shift of the acoustic mode (this mode closely resembles the `radial breathing mode' observed in perfectly circular wires \cite{SINW_111}) as the wire cross-section size is reduced (Fig. \ref{fig:ac_shift} a). There is a clear cross-section shape dependence of the blue shifts where the blue shift follows the order $\Delta \omega_{tri}\;<\Delta \omega_{sq}\;<\Delta \omega_{hex}\;< \Delta \omega_{cir}$. 
The power exponent ($d_{ac}$) varies between 0.6 and 0.856 (Table \ref{table_ac_ph}). The order of the pre-factor ($A_{ac}$) in Eq. (\ref{phonon_shift}) value (Table \ref{table_ac_ph}) also shows the order of the acoustic blue shift for $[$100$]$ SiNWs.

With the increase in wire cross-section size two important things happen, (i) the geometric confinement on the phonons reduces, and (ii) the vibrational energy deviation between the surface and the inner atoms decreases (Fig. \ref{fig:ac_shift}. First point explains the reduction in the blue shift with increase in cross-section size for all the SiNWs while the second point explains the order of blue-shift with shape.

As the SiNW cross-section is reduced lots of low velocity phonon sub-bands start to appear above the acoustic branches (Fig. \ref{fig:SiNW_phonon}) which are normally absent in bulk. These relatively flat bands cause phonon confinement. These zone center low energy phonon sub-bands cause acoustic blue shift. The surface atom vibrations become comparable to the inner core atom vibrations with size reduction. The blue shift is more pronounced when surface modes are less dominant. If the surface atoms vibrations are more than inner atoms then  blue shift is suppressed \cite{raman_size_dep}.

The acoustic blue shift order can be explained by the difference between the vibrational energy density \cite{li_hollow_sinw} of the surface atoms and the inner atoms ($\Delta$E). Surface atoms are the ones with dangling bonds. The spatial vibrational energy density can be calculated as \cite{li_hollow_sinw},

\begin{equation}
\label{eq_eden_phon}
	E^{den}_{i,q} = \sum_{NA}\sum_{j\in[x,y,z]}[F_{BE}(\omega_{i,q})+0.5]\hbar\omega_{i,q}\phi_{i,q,j}\phi^{*}_{i,q,j},
\end{equation}
where NA is the total number of atoms in the unit cell, `i' is the phonon sub-band, `q' is the phonon wave-vector, `j' is the polarization, $F_{BE}$ is Bose-Einstein distribution for phonon population, $\omega_{i,q}$ is the phonon eigen frequency for sub-band `i' and wave-vector `q' and $\phi_{i,q,j}$ is the phonon eigen vibration mode.

\begin{figure}[!htb]
	\centering
		\includegraphics[width=3.3in,height=2.4in]{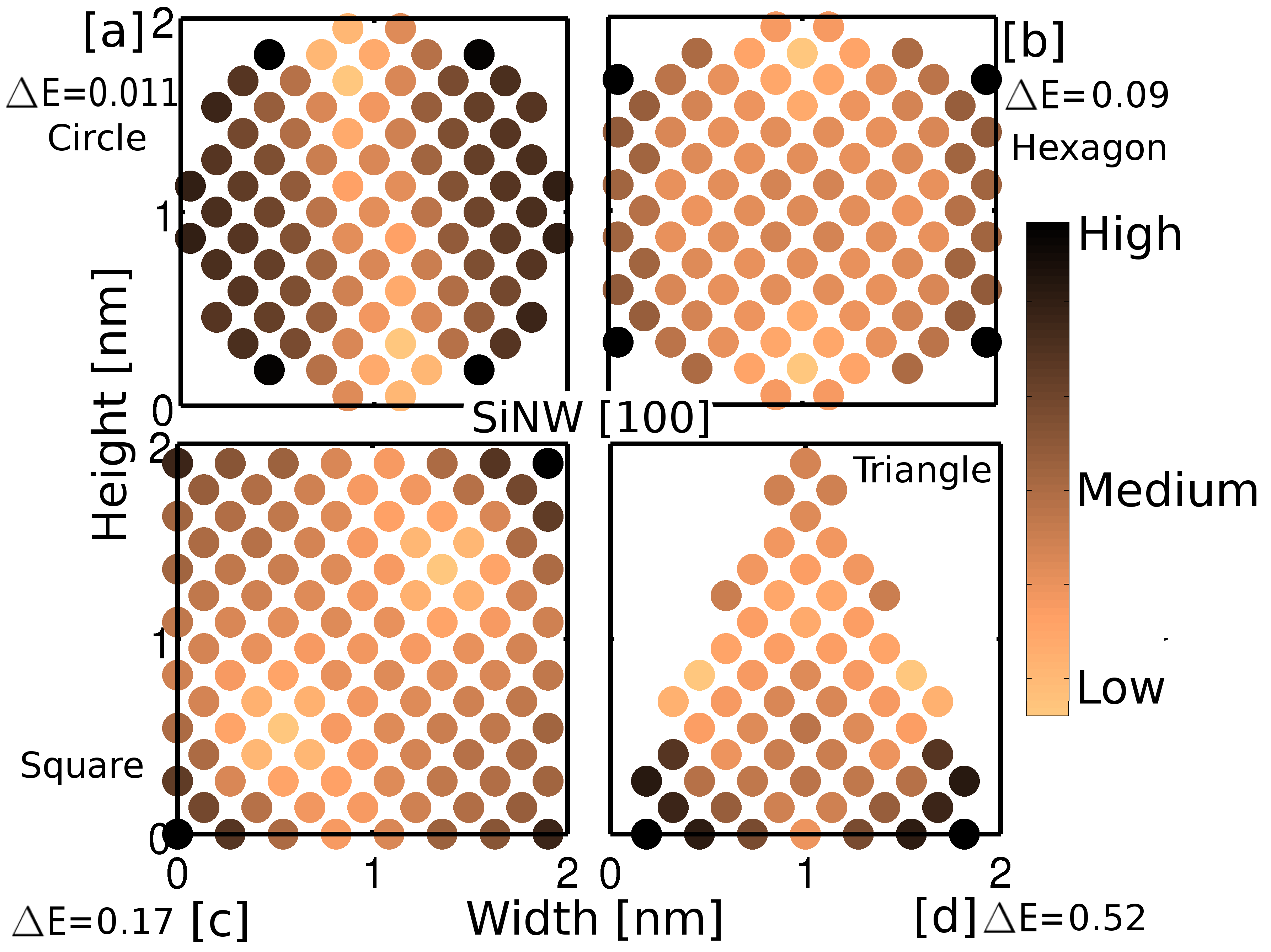} 
	\caption{Phonon energy density for the confined acoustic mode in  2nm $\times$ 2nm $[$100$]$ SiNWs with (a) circular, (b) hexagonal, (c) square and (d) triangular shapes. Surface atoms vibrate more compared to the internal atoms. The difference in the energy of the surface and inner atoms is represented by $\Delta$E. Triangular wires have largest $\Delta$E which results in minimum acoustic phonon confinement (see Fig. \ref{fig:ac_shift}).  }
	\label{fig:ac_shift_eden}
\end{figure}

Figure \ref{fig:ac_shift} b shows that the triangular wire exhibits the maximum difference between the surface and inner atom vibrational energy density for W = 2nm and 4nm, followed by the square, the hexagonal, and the circular wire. Figure \ref{fig:ac_shift_eden} shows the spatial distribution of the energy density for the acoustic phonon branch for 2nm cross-section SiNWs. The surface atoms vibrate more intensively compared to the inner atoms. In circular and hexagonal nanowires all atoms vibrate with a similar intensity as indicated by the small $\Delta$E value. However, in square and triangular wires  the surface atoms vibrate more compared to the inner atoms leading to a larger $\Delta$E value.

The low frequency acoustic phonon modes are very similar to the breathing mode found in larger circular wires \cite{mahan_111}. These eigen frequencies follow a $1/W^{d}$ law (Fig. \ref{fig:ac_shift}), with d = 1 for a perfectly concentric circular wire \cite{mahan_111}. The value of `d' deviates from 1 as the cross-section shape deviates from the ideal circular geometry. 


\subsection{Optical Phonon shift in SiNWs}
\label{sec:Res_2}

\begin{figure}[!htb]
	\centering
		\includegraphics[width=3.0in,height=2.2in]{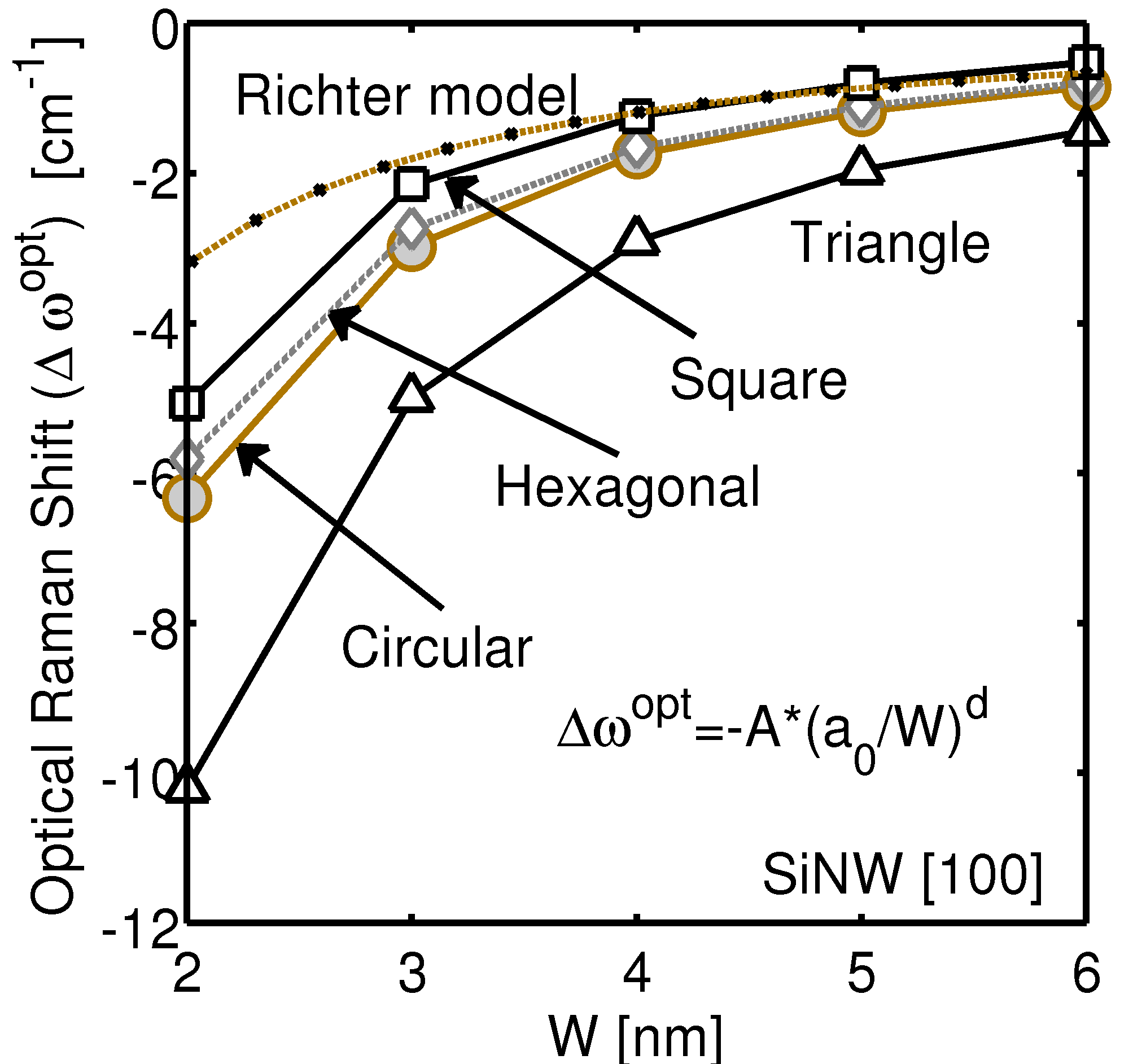}%
	\caption{Effect of cross-section shape on the optical phonon shift in $[$100$]$ SiNWs. }
	\label{fig:opt_shift}
\end{figure}

\textit{Influence of cross-section shape and size}: For optical phonon shift the MVFF model predicts a width exponent factor $d_{opt}$ between 1.8 and 2  (Table \ref{table_1}) for all the cross-section shapes. The $[$100$]$ triangular SiNWs show the maximum optical red shift ($d_{opt}=1.795$) whereas the square SiNWs show the minimum optical red shift ($d_{opt}=2.06$, close to the prediction in Ref.\cite{hepplestone_sinw_2}) (Fig. \ref{fig:opt_shift}). 
Furthermore, with decreasing cross-section size (W) all the wires show an increasing optical red-shift (Fig. \ref{fig:opt_shift}).

Many explanations have been reported in the literature for the optical red-shift in nanostructures like, the increasing surface-to-volume ratio (SVR) \cite{raman_size_dep,size_dep_red_shift}, compressive strain \cite{adv_RS_method}, surface defects \cite{surface_disorder} and optical phonon confinement \cite{size_dep_red_shift,Richter,Campbell,continuum_model}. In the present study of free-standing SiNWs the observed shape and size dependence can be explained using the vibrational energy density \cite{li_hollow_sinw} of the optical mode as well as the optical phonon group velocity ($V^{ph}_{grp}$) \cite{iwce_own_paper}.  

\begin{figure}[!hbt]
	\centering
		\includegraphics[width=2.8in,height=1.9in]{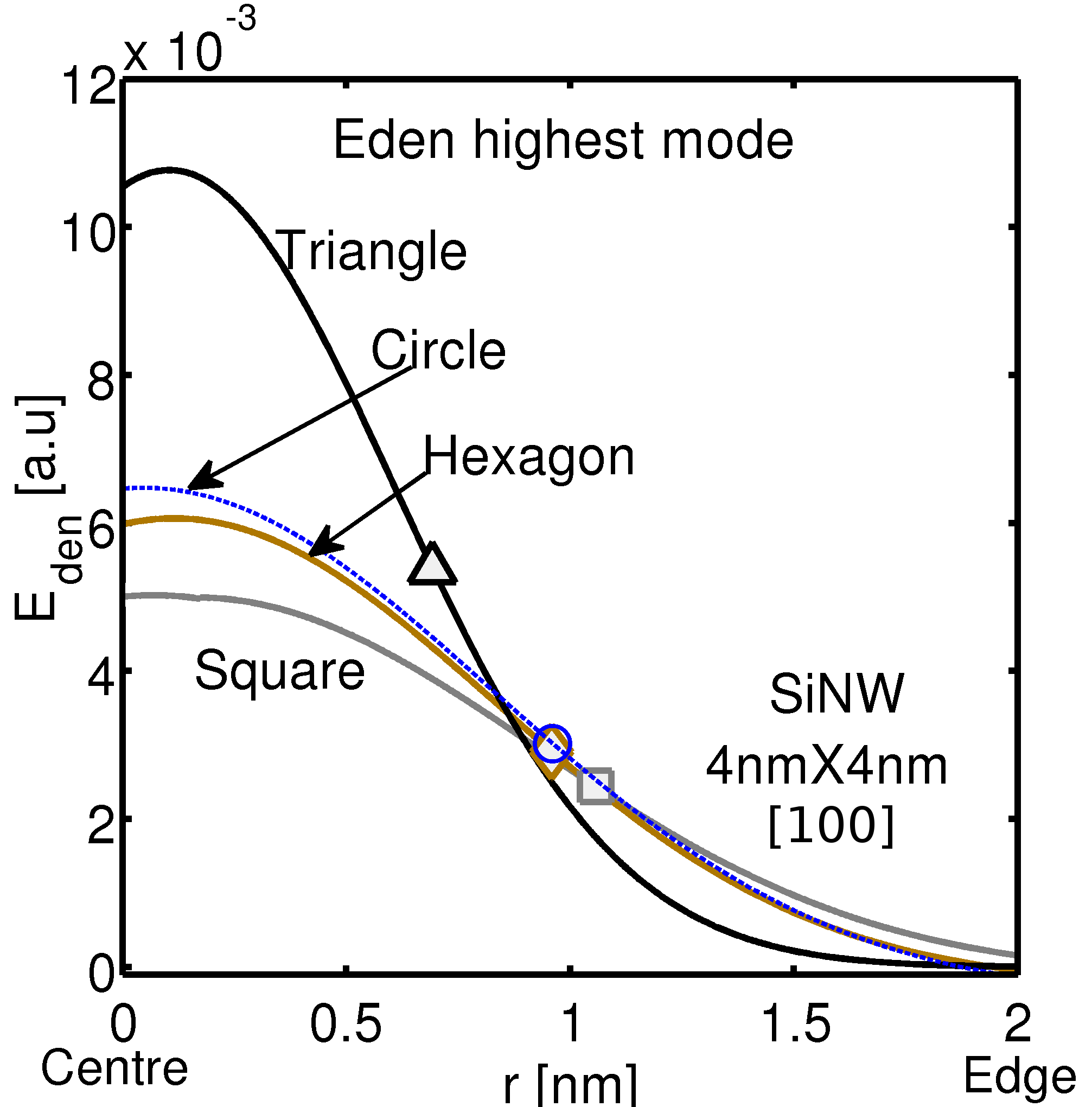}%
	\caption{ Spatial vibrational energy density for optical modes in different cross-section shaped 4nm $\times$ 4nm $[$100$]$ SiNWs.
	The symbols show the full width half maximum (FWHM) of energy density which reflect how active the optical mode is. A larger spatial spread indicates
	more active optical modes. Optical modes in square wires are the closest to the bulk ones and triangular wires are the farthest. }
	\label{fig:opt_Eden}
\end{figure}

Figure \ref{fig:opt_Eden} shows the spatial distribution of the vibrational energy density for the optical mode in SiNWs with W = 4nm. The structure with the maximum spatial spread of the energy density has the most active optical modes \cite{raman_size_dep} since surface atom vibrations contribute to the optical modes. The symbols in Fig. \ref{fig:opt_Eden} show the FWHM extent of the energy density. The square wire has the maximum energy spread while the triangular wire has the minimum energy spread which explains the highest red-shift in the triangular wires (least active optical modes). As the SiNW dimensions decrease, the energy spread also decreases which increases the red shift for all the shapes.

\begin{figure}[!hbt]
	\centering
		\includegraphics[width=2.8in,height=1.8in]{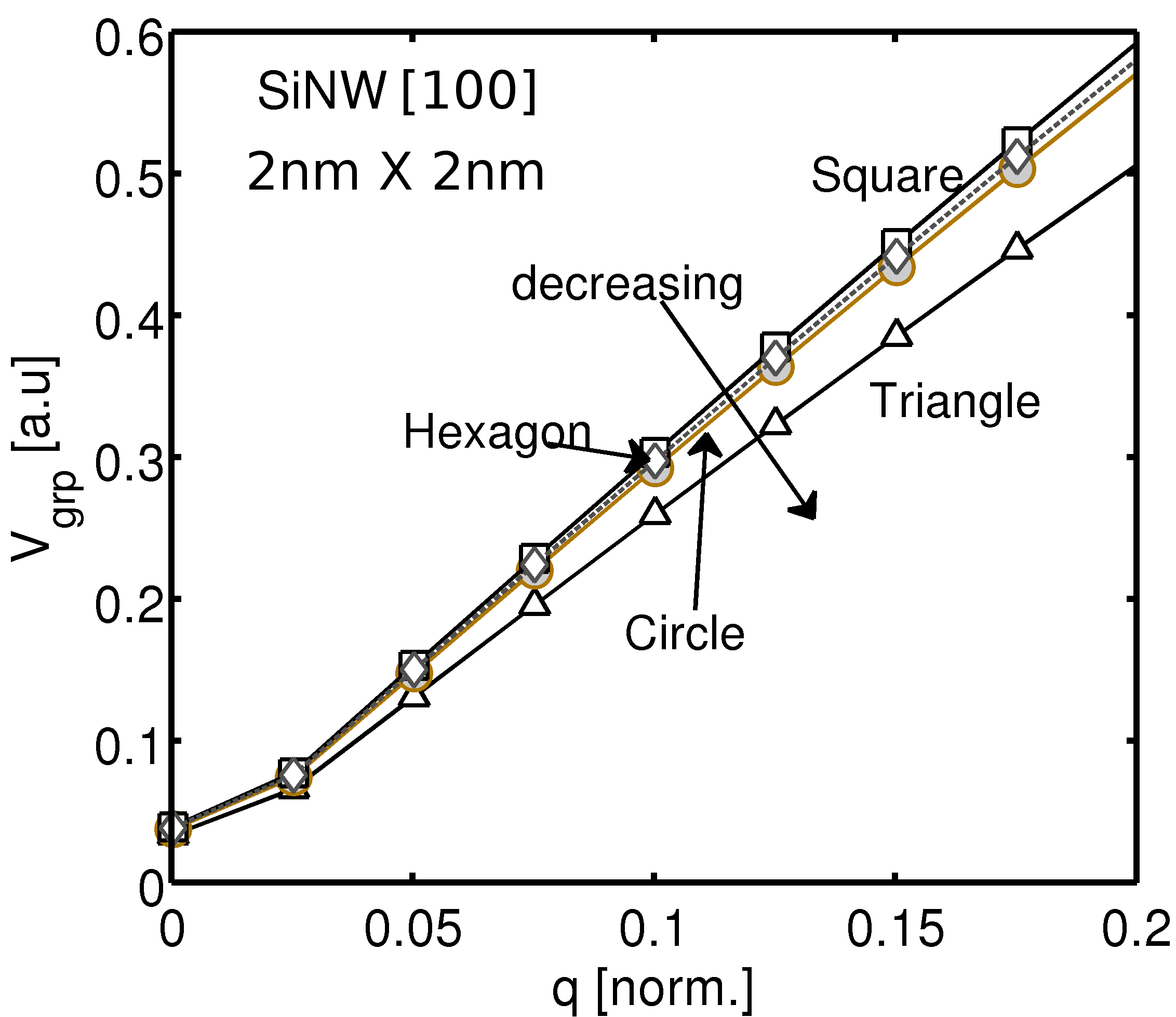}
	\caption{Optical phonon group velocity in $[$100$]$ SiNWs with 2nm $\times$ 2nm cross-section. }
	\label{fig:opt_vgrp}
\end{figure}

Another reason for the red-shift can be attributed to the phonon group velocity. Figure \ref{fig:opt_vgrp} shows that the triangular wire has the smallest group velocity whereas the square wire has the highest group velocity. A smaller group velocity implies higher phonon confinement \cite{continuum_model}. Thus, the phonon group velocity gives the following order of optical phonon confinement; triangular $>$ circular $>$ hexagonal $>$ square. Higher phonon confinement results in more red-shift. This again explains the observed shape dependence of optical red-shift in $[$100$]$ SiNWs.

\subsection{Orientation effect on phonon shifts}
\label{sec:or_dep}

\begin{figure}[!hbt]
	\centering
		\includegraphics[width=3.3in,height=1.7in]{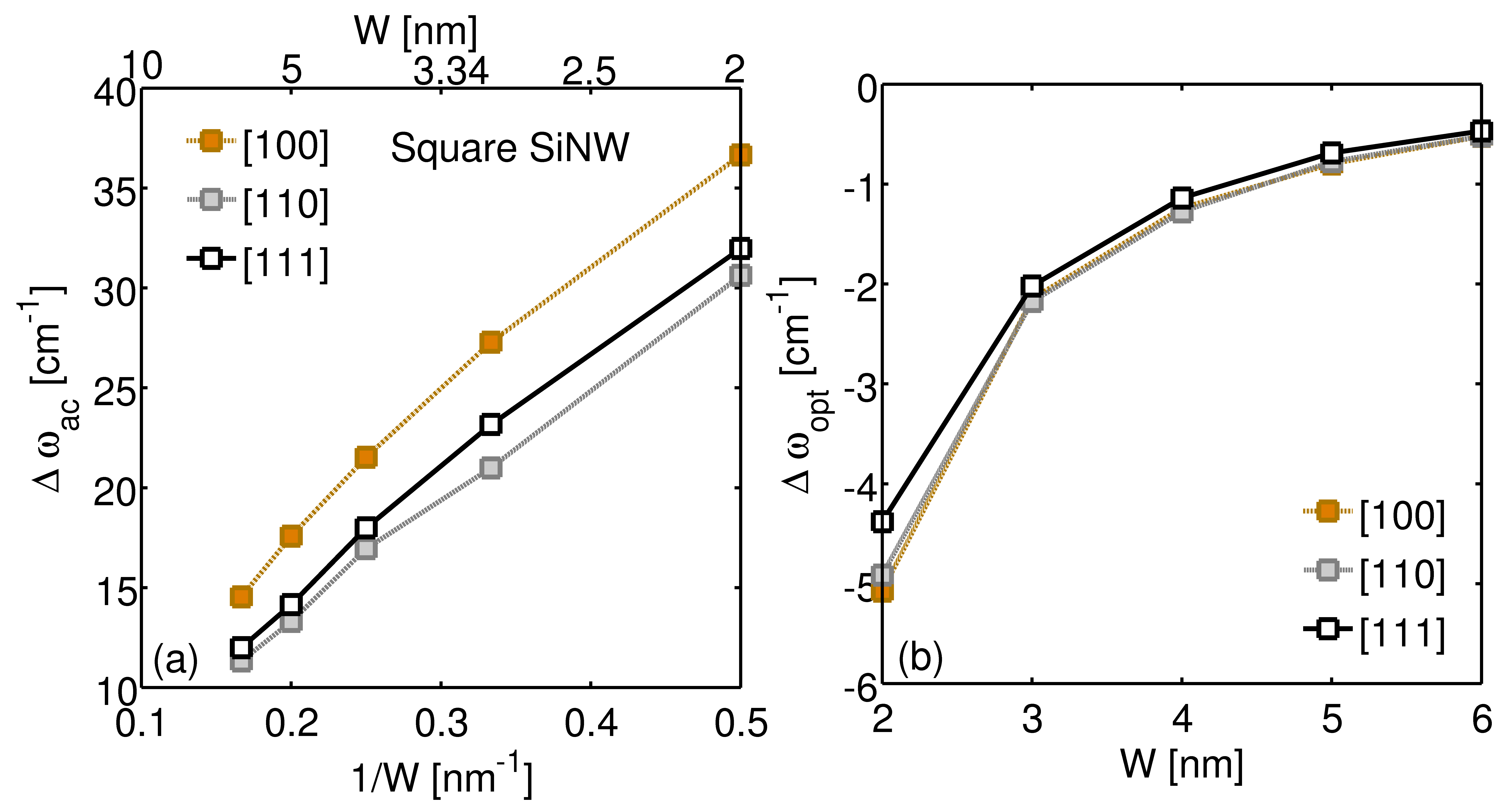}
	\caption{Effect of SiNW orientation on (a) acoustic phonon shift and (b) optical phonon shift. }
	\label{fig:phon_shift_or}
\end{figure}

Along with the shape and size of the SiNWs, the wire axis direction can also produce different phonon shifts. This indicates that the confined phonon dispersion is anisotropic.  For brevity we present the orientation effects for square SiNWs only, however, these trends are also valid for other shapes. The acoustic phonon shift and the optical phonon shifts are shown in Fig. \ref{fig:phon_shift_or} a and b, respectively. The acoustic phonon blue shift exhibits a strong anisotropy (Fig. \ref{fig:phon_shift_or} a), however, the optical red shift has a very weak orientation dependence (\ref{fig:phon_shift_or} b). This observed behavior of phonon shift can be explained by the phonon dispersion anisotropy in Si. In Si the maximum phonon frequency of the acoustic branch along [100], [110] and [111] directions are highly anisotropic \cite{VFF_mod_herman,jauho_method}, whereas the maximum frequency of the optical branches are isotropic ($\sim$519.3 $cm^{-1}$). The acoustic blue shift show the following behavior $\Delta \omega_{ac}^{100} \; > \Delta \omega_{ac}^{111} \; > \Delta \omega_{ac}^{110}$. The cross-section size dependence for both kind of phonon shifts are provided in Table. \ref{table_ornt}.  The acoustic phonon shift exponent of [110] wires is 0.89 which compares reasonably to the exponent of 0.95 observed for [110] SiNWs in  Ref. \cite{SINW_110_phonon} (see Fig. \ref{fig:exp_act_benchmark}).

\subsection{Discussion}

The phonon spectrum shift in SiNWs with shape, size and orientation is useful in understanding the experimental Raman shifts. The analytical expressions for the shifts can provide a good idea about the structure of the SiNWs which can be helpful in determining properties such as phonon-photon interactions. The evolution of the phonon shifts with increasing size also reflects the size (Wlim) at which the shape effects may be no more distinguishable using Raman spectroscopy. From the expresssions of the phonon shifts (Eq. \ref{phonon_shift}) the Wlim value for acoustic blue shift comes out to be roughly 7nm and for optical shifts this value is 10nm. The analytical expressions are well suited for calculating the vibrational energy shifts depending on different physical aspects of the SiNWs.

\section{Conclusion}
\label{sec:conc}

We have used a MVFF model to calculate the acoustic and optical phonon shifts in SiNWs. The theory agrees quite  well with the experimental Raman shift data. At the nanometer scale the acoustic phonon shift is quite sensitive to the (i) wire cross-section size and shape, and (ii) wire channel orientation whereas the optical phonon shift is mainly controlled by the wire cross-section shape and size. The different width exponents and the pre-factors from the acoustic and optical phonon shifts can become a useful tool in predicting the size and shape of the SiNWs. Triangular wires show the highest optical phonon confinement and the least structural symmetry. Analytical expressions for the shape and size dependence of the phonon shifts are useful for compact modeling of physical properties of ultra-scaled SiNWs. 


\section*{Acknowledgment}

The authors acknowledge financial support from MSD Focus Center, under the Focus Center Research Program (FCRP), a Semiconductor Research Corporation (SRC) entity, Nanoelectronics Research Initiative (NRI) through the Midwest Institute for Nanoelectronics Discovery (MIND), NSF (Grant No. OCI-0749140) and Purdue University. Computational support from nanoHUB.org, an NCN operated and NSF (Grant No. EEC-0228390) funded project is also gratefully acknowledged.

%

\newpage
\textbf{TABLES} \\

\begin{table}[htb!]
\centering
\caption{Cross-section shape dependence of longitudinal and transverse sound velocity in W= 3nm, $[$100$]$ SiNW}
\label{table_0}
\begin{tabular}{|c|c|c|}
\hline
Shape & $V_{snd,l} (km/sec)$ & $V_{snd,t} (km/sec)$ \\
\hline
Circular & 6.23 & 4.50\\
Hexagon & 6.18 &  4.37\\
Square &  6.26 &  4.37\\
Triangular & 5.79 & 3.12\\
\hline
\end{tabular}
\end{table}

\begin{table}[htb!]
\centering
\caption{Width parameters for acoustic phonon shift in $[$100$]$ SiNWs}
\label{table_ac_ph}
\begin{tabular}{|l|c|c|}\hline
Shape & $A_{ac}\;(cm^{-1})$  & $d_{ac}$ \\
\hline
Circular & 140.05 & 0.856 \\
Hexagon & 117.63 & 0.824 \\
Square & 106.37 & 0.81 \\
Triangular & 59.20 & 0.6  \\
\hline
\end{tabular}
\end{table}

\begin{table}[!htb]
\centering
\caption{Width parameters for optical phonon shift in $[$100$]$ SiNWs}
\label{table_1}
\begin{tabular}{|l|c|c|}\hline
Shape & $A_{opt}\;(cm^{-1})$& $d_{opt}$\\
\hline
Circular &-70.49 & 1.85 \\
Hexagon &-62.59 & 1.824\\
Square & -74.47& 2.06\\
Triangular & -106.05 &1.795 \\
\hline
\end{tabular}
\end{table}

\begin{table}[!htb]
\centering
\caption{Width parameters for acoustic and optical phonon shift in square SiNWs with different orientations}
\label{table_ornt}
\begin{tabular}{|c|c|c|c|c|}\hline
Channel Or	& $A_{ac}\;(cm^{-1})$	& $d_{ac}$ 	& $A_{opt}\;(cm^{-1})$ 	& $d_{op}$\\
\hline
$[$100$]$ 		&  106 		& 0.81 		&	-74.62 	& 2.06  \\
$[$110$]$ 		&  98.34 	& 0.89 		& -66.47 	& 2.00 \\
$[$111$]$ 		& 	100.7	& 0.87 		& -57.17		& 1.97 \\
\hline
\end{tabular}
\end{table}

\newpage%
\newpage
\textbf{FIGURE TITLES} \\

\end{document}